\documentclass[aps,pra,twocolumn,superscriptaddress,showpacs,floatfix]{revtex4}
\usepackage{graphicx}

\begin{document}

\title{Faraday waves in collisionally inhomogeneous Bose-Einstein condensates}

\author{Antun Bala\v{z}}
\email{antun@ipb.ac.rs}
\affiliation{Scientific Computing Laboratory, Institute of Physics Belgrade,
University of Belgrade, Pregrevica 118, 11080 Belgrade, Serbia}

\author{Remus Paun}
\affiliation{Horia Hulubei National Institute of Physics and Nuclear Engineering (IFIN-HH), Department of Computational
Physics and Information Technologies, P.~O.~B. MG-6, 077125, Romania}

\author{Alexandru I. Nicolin}
\affiliation{Horia Hulubei National Institute of Physics and Nuclear Engineering (IFIN-HH), Department of Computational
Physics and Information Technologies, P.~O.~B. MG-6, 077125, Romania}

\author{Sudharsan Balasubramanian}
\affiliation{Centre for Nonlinear Science, Post-Graduate and Research Department
of Physics, Government College for Women (Autonomous), Kumbakonam 612001, India}

\author{Radha Ramaswamy}
\affiliation{Centre for Nonlinear Science, Post-Graduate and Research Department
of Physics, Government College for Women (Autonomous), Kumbakonam 612001, India}

\begin{abstract}
We study the emergence of Faraday waves in cigar-shaped
collisionally inhomogeneous Bose-Einstein condensates subject to periodic
modulation of the radial confinement. Considering a Gaussian-shaped
radially inhomogeneous scattering length, we show through extensive
numerical simulations and detailed variational treatment that the spatial period
of the emerging Faraday waves increases as the inhomogeneity of the
scattering length gets weaker, and that it saturates
once the width of the radial inhomogeneity reaches the radial
width of the condensate. In the regime of strongly inhomogeneous scattering lengths,
the radial profile of the condensate is akin to that of a hollow cylinder,
while in the weakly inhomogeneous case the condensate is cigar-shaped and has a Thomas-Fermi radial density profile.
Finally, we show that when the frequency of the modulation is close
to the radial frequency of the trap, the condensate exhibits resonant
waves which are accompanied by a clear excitation of collective modes, while for
frequencies close to twice that of the radial frequency of the trap, the observed
Faraday waves set in forcefully and quickly destabilize condensates
with weakly inhomogeneous two-body interactions.
\end{abstract}


\maketitle

\section{Introduction}

Over the past two decades ultracold quantum gases have been an almost perfect
playground for nonlinear scientists due to their versatility
and excellent experimental control.
Bose-Einstein condensates (BECs) have been particularly attractive
\cite{PanosBook,TheBECBook} due to several reasons: the extreme tunability
of their short-range two-body interactions (using either magnetic
or optical Feshbach resonances), the existence of atomic species which
also posses long-range dipole-dipole interactions, the experimental
condensation of multi-component systems (realized with one atomic species in two or more distinct hyperfine states
or with several distinct atomic species) with tunable inter-state or inter-species
interactions, and the possibility to modify the geometry of condensates
almost at will \cite{TheBECBook}. Moreover, this remarkable level
of experimental control was accompanied by an accurate theoretical description at mean-field level,
using the so-called Gross-Pitaevskii equation \cite{OriginalGPE},
which catalyzed thorough investigations into the dynamics of the condensates at (or close to) zero temperature.

Pattern formation in quantum fluids is a related research topic which has been intensively studied
and there are by now experimental results on the
emergence of Faraday patterns in BECs \cite{FWBECExp} and He$^4$
cells \cite{FWHeCellsExp} subject to parametric drives, as well as
numerous theoretical studies on Faraday waves in condensates with
short-range interactions \cite{FWHist,FWHistF}, dipolar condensates
\cite{FWSantos}, binary condensates with short-range interactions
\cite{FWBalaz}, Fermi-Bose mixtures \cite{FWAbdullaev}, and superfluid
Fermi gases \cite{FWFermi}. Moreover, it has been shown that Faraday
waves can be suppressed in condensates subject either to resonant
parametric modulations \cite{RWNicolin} or space- and time-modulated
potentials \cite{FWinOL,CleaningBECs}, which is a widely studied topic
\cite{Gaul1,Gaul2,Schmelcher1,Schmelcher2,Ivana,Kobyakov,Hu,Sakhel,td1}.
Furthermore, in the context of parametric excitations,
the formation of density patterns has been studied in expanding ultra-cold
Bose gases (either fully \cite{SelfInducedDensMod} or only partially
condensed \cite{DensRipplesExpandingBECsTh,DensRipplesExpandingBECsExp}),
and the spontaneous formation of density waves has been reported for
antiferromagnetic BECs \cite{SpontaneousPattForm}. 

In this paper we focus on cigar-shaped condensates with Gaussian-shaped
radially inhomogeneous scattering length subject to periodic modulation
of the radial confinement. Such systems are in the so-called
collisionally inhomogeneous regime \cite{CollisInhom} which can be
achieved either by magnetic or optical means. Magnetic Feshbach resonances
are well-established experimental methods and have been used to study
the formation of ultracold molecules \cite{ColdMoleculesFeshbach},
the BEC-BCS crossover \cite{BECBCSFeshbach}, and the production of
Efimov trimer states \cite{EfimovFeshbach}, but the length scale
for application of the Feshbach field is usually larger than the size
of the atomic BEC sample, so these methods could not be used in reaching
the collisionally inhomogeneous regime. Optical Feshbach resonances,
however, have been shown to allow fine spatial control of the scattering
length and recent experimental results show modulations of the $s$-wave
scattering length on the scale of hundreds of nanometers \cite{OFR}.
Furthermore, it has been shown that the collisionally inhomogeneous
regime supports a plethora of new nonlinear phenomena such as the
adiabatic compression of matter-waves \cite{CollisInhom,AdiabComp},
Bloch oscillations of matter-wave solitons \cite{CollisInhom}, atomic
soliton emission and atom lasers \cite{AtomicSolitonEmission}, dynamical
trapping of matter-wave solitons \cite{DynTrap,DynTrap1,DynTrap2,Delgado,Lee}, enhancement
of transmissivity of matter-waves through barriers \cite{DynTrap1,DynTrap2,EnhancementTrans},
formation of stable condensates exhibiting both attractive and repulsive
interatomic interactions \cite{AttRep,Adhikari1,Garcia-March}, the delocalization transition
in optical lattices \cite{DelocTrans}, spontaneous symmetry breaking
in a nonlinear double-well pseudo-potential \cite{NonlinDW}, the
competition between incommensurable linear and nonlinear lattices
\cite{DynTrap2,LinearNonlinLatt}, the generation of solitons \cite{GenDarkSolitons,Dey}
and vortex rings \cite{VortexRings}, and many others. 

Here we show through extensive numerical simulations and 
supporting variational calculations that the spatial period of the Faraday waves which 
emerge in collisionally inhomogeneous condensates depends strongly
on the space modulation of the scattering length. In particular, we
show that the spatial period increases as the inhomogeneity becomes weaker
and that it saturates once
the width of the Gaussian-shaped inhomogeneity approaches the radial
width of the condensate. As we will show, this behavior can be understood 
in terms of an effective nonlinearity of the system, which reveals that the system 
becomes more nonlinear as the inhomogeneity becomes weaker, thereby 
exhibiting clearly observable Faraday waves of longer spatial periods
and shorter instability onset times.
In the regime of strongly inhomogeneous collisions,
the radial profile of the condensate resembles that of a hollow cylinder,
while in the weakly inhomogeneous case the condensate is cigar-shaped and has a Thomas-Fermi radial density profile.
The latter regime can be described using the usual variational description
of density waves in BECs  \cite{FWHistF}, while for the former we introduce a versatile
trial wave function which describes both the bulk properties of the
condensate and the emergence of the density wave. 

The paper is structured as follows: in Sec.~\ref{sec:var} we introduce
the Gross-Pitaevskii equation and detail the variational treatment
of density waves, and in Sec.~\ref{sec:res} we present our numerical
and analytical results. Finally, Sec.~\ref{sec:con} gathers our concluding
remarks and gives outlook for future research.

\section{Variational treatment of the Gross-Pitaevskii equation}
\label{sec:var}

The ground state properties and the dynamics of a three-dimensional
BEC at zero temperature are accurately described, respectively, by the time-independent 
\begin{equation}
\left(-\frac{\hbar^2}{2m}\Delta+V({\bf r})+g({\bf r}) N \left|\psi\right|^{2}\right)\psi=\mu\psi\, ,
\label{eq:GPEind}
\end{equation}
and time-dependent
\begin{equation}
i\hbar\frac{\partial\psi}{\partial t}=\left(-\frac{\hbar^2}{2m}\Delta+V({\bf r})+g({\bf r},t) N \left|\psi\right|^{2}\right)\psi\, ,
\label{eq:GPEtdep}
\end{equation}
versions of the Gross-Pitaevskii equation (GPE). Here, $\mu$ is chemical potential of the system, $N$ is the total number of atoms in a BEC, and
\begin{equation}
V({\bf r})=\frac{m}{2}\left(\Omega_{\rho}^{2}\rho^{2}+\Omega_{z}^{2}z^{2}\right)
\end{equation}
represents the external confining potential, which may depend on time through frequencies $\Omega_\rho=\Omega_\rho(t)$ and $\Omega_z=\Omega_z(t)$. The
strength of the nonlinear interaction $g$ is proportional
to the $s$-wave scattering length $a_s$,
\begin{equation}
g=\frac{4\pi\hbar^2}{m} a_s\, .
\end{equation}
and can be engineered to be spatially inhomogeneous (e.g., by using optical Feshbach resonances), or time-dependent (e.g., by harmonic modulation of the applied magnetic field close to a Feshbach resonance), or both.

The previous GPE equations can
be solved numerically without difficulty using readily available sequential Fortran
codes \cite{GPENumerics1} or OpenMP-parallelized C codes \cite{GPENumerics2} which implement Crank-Nicolson methods, but other numerical approaches are also available \cite{num1,num2,num3,num4,num5,num6}.
However, for analytical insights into the dynamics of the condensate, such numerical calculations are usually accompanied
by variational or hydrodynamical approaches \cite{TheBECBook}.
Variational methods are particularly attractive because one can simplify
the dynamics of the condensate to a coupled system of ordinary differential
equations from which one can analytically determine the frequencies
of the collective excitations, the speed of sound in the condensate,
the position of resonances, etc. To this end, one starts from the
Gross-Pitaevskii Lagrangian density
\begin{equation}
{\cal L}({\bf r},t)=\frac{\hbar^2}{2m}\left|\nabla\psi\right|^{2}+V({\bf r},t)\left|\psi\right|^{2}+\frac{gN}{2}\left|\psi\right|^{4}\, ,
\label{eq:Ldens}
\end{equation}
which is then minimized for a selected trial wave function that captures
the physics of the problem under scrutiny.

In our case, we consider
a longitudinally homogeneous cigar-shaped condensate, i.e.,
$\Omega_{z}(t)=0$, whose radial frequency is harmonically modulated in time,
\begin{equation}
\Omega_\rho(t)=\Omega_{\rho 0}(1+\epsilon\sin\omega t)\, ,
\end{equation}
where $\epsilon$ is a modulation amplitude, and $\omega$ is a modulation frequency.
Furthermore, the scattering length is spatially modulated in the radial direction such that the nonlinear interaction has the form
\begin{equation}
g=g(\rho)  =  \frac{4\pi\hbar^2 a(0)}{m}\, e^{-\frac{\rho^2}{2b^2}}= g_0\, e^{-\frac{\rho^2}{2b^2}}\, ,
\label{eq:g}
\end{equation}
where $a(0)=a_s|_{\rho=0}$ is the (constant) value of the $s$-wave scattering length along the $z$-axis, and $b$ is the length scale of the space modulation of the scattering length in the radial direction.

The trial wave function that captures the dynamics of collisionally inhomogeneous BEC is chosen as 
\begin{eqnarray}
\psi({\bf r},t) & = & \phi({\bf r},t)\left\{1+[u(t)+iv(t)]\cos kz\right\}\nonumber \\
 & = & A(t)\cdot\left(1+\gamma\rho^{2}\right)\exp\left(-\frac{\rho^{2}}{2w^{2}(t)}+i\rho^{2}\alpha(t)\right)\nonumber \\
  & & \times\left\{1+[u(t)+iv(t)]\cos kz\right\}\, ,
\label{eq:trialfun}
\end{eqnarray}
where $A(t)$ is chosen such that the density is normalized to unity
over one period of $\cos kz$, i.e., over the interval $[-\pi/k, \pi/k]$. Note that the trial wave function
consists of the radial envelope $\phi({\bf r},t)$ that describes the collective dynamics
of the condensate, multiplied by a periodic function that captures the emergence
of longitudinal density waves.
We stress here that effectively one-dimensional systems like the one we are investigating exhibit only one-dimensional patterns (i.e., waves), which all look alike, independently of the spatial inhomogeneity of the scattering length. Therefore, we focus on the study of their spatial periods and the instability onset times. The two- and three-dimensional systems, however, are qualitatively different because in these cases the spatial distribution of the scattering length impacts the geometry of the excited patterns to the extent of having, for instance, transitions from triangular to square patterns after small modifications of the scattering length, and this will be the topic of our forthcoming publication.

To arrive at the desired equations governing the dynamics,
one integrates the Lagrangian density over one spatial period of the density
wave and minimizes the ensuing (time-dependent) Lagrangian through
the classical Euler-Lagrange equations \cite{GaussVar}, which yield
in our case four ordinary differential equations that correspond to
minimizations with respect to variational parameters $w(t)$, $\alpha(t)$, $u(t)$ and
$v(t)$, and one algebraic equation that corresponds to minimization
with respect to the parameter $\gamma$. The physical interpretation of the variational parameters is quite simple:
$w(t)$ corresponds to the width of the condensate, $\alpha(t)$ is the corresponding phase, and $u(t)+iv(t)$ is the complex amplitude of the density wave, while $\gamma$ is less transparent and measures the inhomogeneity of the collisions. The spatial period of the grafted wave, i.e., $2\pi/k$, is determined by considering the necessary conditions for the emergence of density waves, therefore $k$ is not treated here as a variational parameter. The quality of the variational results
depends strongly on how accurately the trial wave function describes
the possible modes of the condensates, and numerous
other options are explored in the literature (see Refs.~\cite{VarCollectiveModes,VariationalDensityWaves,VariationalDensityWavesF} 
and references therein for the main results). 

Finally, let us also notice that improved accuracy usually comes at the
cost of cumbersome variational equations which are hard to investigate
by purely analytical means. Consequently, instead of the general set
of equations which describes both the weakly- and strongly-inhomogeneous
regime, we focus in the next subsections on two distinct
simplified sets of equations, one for each regime.
The good agreement with the numerical results, presented in Sec.~\ref{sec:res}, fully justifies the use of the variational trial function (\ref{eq:trialfun}).
For simplicity,  from now on we will use the natural system of units ($\hbar=m=1$).

\subsection{Weakly inhomogeneous collisions}
\label{sec:varweak}

The regime of weakly inhomogeneous collisions corresponds to large
values of the length scale $b$, such that the exponential term in Eq.~(\ref{eq:g})
is very close to unity. The stationary density profile of the condensate 
obtained by numerically solving Eq.~(\ref{eq:GPEind}) shows
a strong localization of the atoms around the symmetry axis of the
condensate, and one can safely investigate its dynamics considering
variational parameter $\gamma$ to be small. Within this approximation we have the following variational equations for the dynamics of the bulk of the condensate:
\begin{eqnarray}
\gamma & = & \frac{4b^{2}+\tilde{w}^{2}}{4\tilde{w}^{2}\left(8b^{6}E_{1}+16b^{4}E_{2}\tilde{w}^{2}+4b^{2}E_{2}\tilde{w}^{4}+\pi\tilde{w}^{6}\right)}\times\nonumber \\
 &  & \times\left\{ 8b^{4}E_{3}+4b^{2}E_{3}\tilde{w}^{2}+\pi\left(1-16b^{4}\Omega_{\rho 0}^{2}\right)\tilde{w}^{4}\right.\nonumber \\
 &  & \left.+8\pi b^{2}\Omega_{\rho 0}^{2}\tilde{w}^{6}+\pi\Omega_{\rho 0}^{2}\tilde{w}^{8}\right\} ,\label{eq:wicgamma}
\end{eqnarray}
\begin{eqnarray}
\ddot{w}(t) & = & \frac{2\pi+ng_0}{2 \pi w(t)^{3}}-\frac{2\gamma\left(2\pi+ng_0\right)}{\pi w(t)}\nonumber \\
 &  & -\frac{w(t)}{2\pi\left(4b^{2}+w(t)^{2}\right)^{3}}\Big\{ 4b^{2}\left(ng_0+32\pi b^{4}\Omega_{\rho}(t)^{2}\right)\nonumber \\
 &  & +2\pi\Omega_{\rho}(t)^{2}w(t)^{6}+\left[ng_0+96\pi b^{4}\Omega_{\rho}(t)^{2}\right] w(t)^{2}\nonumber \\
 &  & -4\left[\gamma ng_0-6\pi b^{2}\Omega_{\rho}(t)^{2}\right] w(t)^{4}\Big\} .\label{eq:wicw}
\end{eqnarray}
For the density wave, the variational equations have the form:
\begin{eqnarray}
\dot{u}(t) & = & \frac{k^{2}v(t)}{2}\, ,\label{eq:wicu}\\
\dot{v}(t) & = & -\left(\frac{k^{2}}{2}+\frac{4b^{2}ng_0}{\pi w(t)^{2}\left(4b^{2}+w(t)^{2}\right)}\right)u(t)\, .\label{eq:wicv}
\end{eqnarray}
In previous equations, $\tilde{w}$ is the equilibrium width
of the condensate obtained from Eq.~(\ref{eq:wicw}) with $\epsilon=0$ (i.e., $\Omega_\rho=\Omega_{\rho 0}$),  $E_{1}=8\pi+3ng_0$, and $E_{2}=3\pi+ng_0$, $E_{3}=2\pi+ng_0$, while
$n$ is the longitudinal density of the condensate. The above equations represent
the truncated version of the full set of Euler-Lagrange equations obtained by neglecting 
terms of the order ${\cal O}\left(\gamma^{2}\right)$.

The important point for our analysis is that Eqs.~(\ref{eq:wicu})-(\ref{eq:wicv}) 
can be cast into a Mathieu-like equation where the parametric drive is due to
the time dependence of $w(t)$, namely
\begin{equation}
\ddot{u}(\tau)+u(\tau)[A_\mathrm{W}(k,\omega)+\epsilon B_\mathrm{W}(k,\omega) \sin 2\tau]=0\, ,\label{eq:MathieuNonlinearA}
\end{equation}
where $\omega t=2\tau$, and the coefficient $A_\mathrm{W}(k,\omega)$,
\begin{equation}
A_\mathrm{W}(k, \omega)=\frac{2 k^{2}}{\omega^2}\left(\frac{k^{2}}{2}+\frac{4b^{2}ng_0}{\pi\tilde{w}^{2}\left(4b^{2}+\tilde{w}^{2}\right)}\right)\, ,
\end{equation}
will be relevant for calculation of the spatial period of Faraday patterns.
The density waves described by this equation emerge due to
the periodic modulation of the strength of the confining potential,
which in turn generates periodic oscillations of the radial width of the
condensate $w(t)$, and thereby serves as an effective parametric
drive in Eq.~(\ref{eq:MathieuNonlinearA}).

General solutions of Eq.~(\ref{eq:MathieuNonlinearA}) are not known analytically,
but for small values of the modulation amplitude $\epsilon$ the equation reduces to a pure
Mathieu equation whose solutions are well-known \cite{MathieuBook}.
The Faraday waves observed experimentally \cite{FWBECExp,FWHeCellsExp} correspond to the most unstable
solutions, which are first excited, and their dispersion relation $k(\omega)$ is obtained from
the condition $A_\mathrm{W}(k,\omega)=1$ \cite{MathieuBook}, which yields
\begin{eqnarray}
k_\mathrm{F,W}&=&\Bigg\{ \sqrt{\omega^{2}+\frac{16b^{4}n^2g_0^{2}}{\pi^{2}\tilde{w}^{4}\left(4b^{2}+\tilde{w}^{2}\right)^{2}}}\nonumber\\
&&-\frac{4b^{2}ng_0}{\pi\tilde{w}^{2}\left(4b^{2}+\tilde{w}^{2}\right)}\Bigg\} ^\frac{1}{2}\, .
\label{eq:dispersionA}
\end{eqnarray}
From this expression, the spatial period of Faraday waves is calculated as $p=2\pi/k_\mathrm{F,W}$.
 
\subsection{Strongly inhomogeneous collisions}
\label{sec:varstrong}

The regime of strongly inhomogeneous collisions, i.e., strongly spatially modulated interactions corresponds to small
values of the parameter $b$. It is qualitatively different from the regime of weakly inhomogeneous collisions
as the condensate has a stationary density profile akin to that of a hollow cylinder. This is due to the fact that the interaction energy decreases as the condensate is further away from the longitudinal axis, where it has a maximum, while the potential energy increases with the increase of the radial distance from the longitudinal axis. Therefore, the ground state, which has a minimal total energy, is found in-between, with the maximal density of the condensate at some distance from the longitudinal axis, depending on the strength of inhomogeneity. 

To describe analytically the condensate in the regime of strong inhomogeneity, we
consider $\gamma$ to be large, such that the corresponding Euler-Lagrange
equations for the bulk of the condensate can be truncated to  
\begin{eqnarray}
\gamma & = & \frac{2}{\tilde{w}^{2}}\left\{ \frac{1024b^{8}}{\tilde{w}^{2}}+1280b^{6}+160b^{2}\tilde{w}^{4}\right.\nonumber \\
 &  & +\left.8b^{4}\left(80+\frac{3ng_0}{\pi}\right)\tilde{w}^{2}+20\tilde{w}^{6}+\frac{\tilde{w}^{8}}{b^{2}}\right\} \times\nonumber \\
 &  & \times\Big\{ \frac{1024b^{8}}{\tilde{w}^{2}}+32b^{6}\left(40-\frac{3ng_0}{\pi}\right)+\frac{\tilde{w}^{8}}{b^{2}}\nonumber \\
 &  & +128b^{4}\left(5+8b^{4}\Omega_{\rho 0}^{2}\right)\tilde{w}^{2}+\frac{\Omega_{\rho 0}^{2}\tilde{w}^{12}}{b^{2}}\nonumber \\
 &  & +160b^{2}\left(1+\Omega_{\rho 0}^{2}\tilde{w}^{4}+8b^{4}\Omega_{\rho 0}^{2}\right)\tilde{w}^{4}\nonumber \\
 &  & +20\left(1+\Omega_{\rho 0}^{2}\tilde{w}^{4}+32b^{4}\Omega_{\rho 0}^{2}\right)\tilde{w}^{6}\Big\}\, ,\label{eq:sicgamma}
\end{eqnarray}
\begin{eqnarray}
\ddot{w}(t) & = & \frac{1}{3w(t)^{3}}-\frac{2}{3\gamma w(t)^{5}}-\Omega_{\rho}(t)^{2}w(t)+\frac{256b^{12}ng_0}{\pi E_4 w(t)^3}\nonumber \\
 &  & +\frac{128b^{10}ng_0}{\pi\gamma E_4 w(t)^3}+\frac{384b^{10}ng_0}{\pi E_4 w(t)}+\frac{192b^{8}ng_0}{\pi\gamma E_4 w(t)}\, ,\label{eq:sicw}
\end{eqnarray}
while for the density wave the equations are truncated to
\begin{eqnarray}
\dot{u}(t) & = & \frac{k^{2}}{2}v(t)\, ,\label{eq:sicu}\\
\dot{v}(t) & = & -\left(\frac{k^{2}}{2}+\frac{384b^{10}ng_0}{\pi w(t)^{2}\left(4b^{2}+w(t)^{2}\right)^{5}}\right)u(t)\, ,\label{eq:sicv}
\end{eqnarray}
where $E_{4}=\left(4b^{2}+w(t)^{2}\right)^{6}$.
Equation (\ref{eq:sicgamma}) is correct up to terms of the order ${\cal O}\left(\gamma^{-4}\right)$,
while the other equations are correct to to terms of the order ${\cal O}\left(\gamma^{-2}\right)$.
As before, $n$ represents the  longitudinal, radially-integrated density of the condensate.

The next crucial step is to cast the last two equations into a single Mathieu-like equation,
\begin{equation}
\ddot{u}(\tau)+u(\tau)[A_\mathrm{S}(k,\omega)+\epsilon B_\mathrm{S}(k,\omega) \sin 2\tau]=0\, ,
\label{eq:MathieuNonlinearB}
\end{equation}
where again  $\omega t=2 \tau$, and the coefficient $A_\mathrm{S}(k,\omega)$ has the form
\begin{equation}
A_\mathrm{S}(k,\omega)=\frac{2k^2}{\omega^2}\left(\frac{k^2}{2}+\frac{384b^{10}ng_0}{\pi \tilde{w}^2
\left(4b^2+\tilde{w}^2\right)^5}\right)\, .
\end{equation}

The most unstable solution of the above Mathieu equation is again given by the condition $A_\mathrm{S}(k,\omega)=1$, 
yielding
\begin{eqnarray}
\hspace*{-5mm}
k_\mathrm{F,S} & = & \frac{1}{\sqrt{\pi\tilde{w}^{2}\left(4b^{2}+\tilde{w}^{2}\right)^{5}}}\Bigg\{ -384b^{10}ng_0\nonumber \\
 &  & +\sqrt{Cb^{20}n^2g_0^2+\pi^{2}\tilde{w}^{4}\left(4b^{2}+\tilde{w}^{2}\right)^{10}\omega^{2}}\Bigg\} ^\frac{1}{2}\, ,
\label{eq:dispersionB}
\end{eqnarray}
where constant $C$ has a value $C=147,456$ and $\tilde{w}$ is the equilibrium width of the
condensate obtained from Eq.~(\ref{eq:sicw}) for $\epsilon=0$ (i.e., $\Omega_\rho=\Omega_{\rho 0}$). As before, the spatial period of emerging Faraday waves is calculated as $p=2\pi/k_\mathrm{F,S}$.

The dispersion relation for the case of weak inhomogeneity, Eq.~(\ref{eq:dispersionA}), and for the case of strong inhomogeneity,
Eq.~(\ref{eq:dispersionB}), represent main contributions of this paper,
and we show in the next section that they accurately describe
the properties of density waves in realistic condensates, despite
the simplifications that have been used in the variational approach.

\section{Results}
\label{sec:res}

\begin{figure*}
\includegraphics[height=5.9cm]{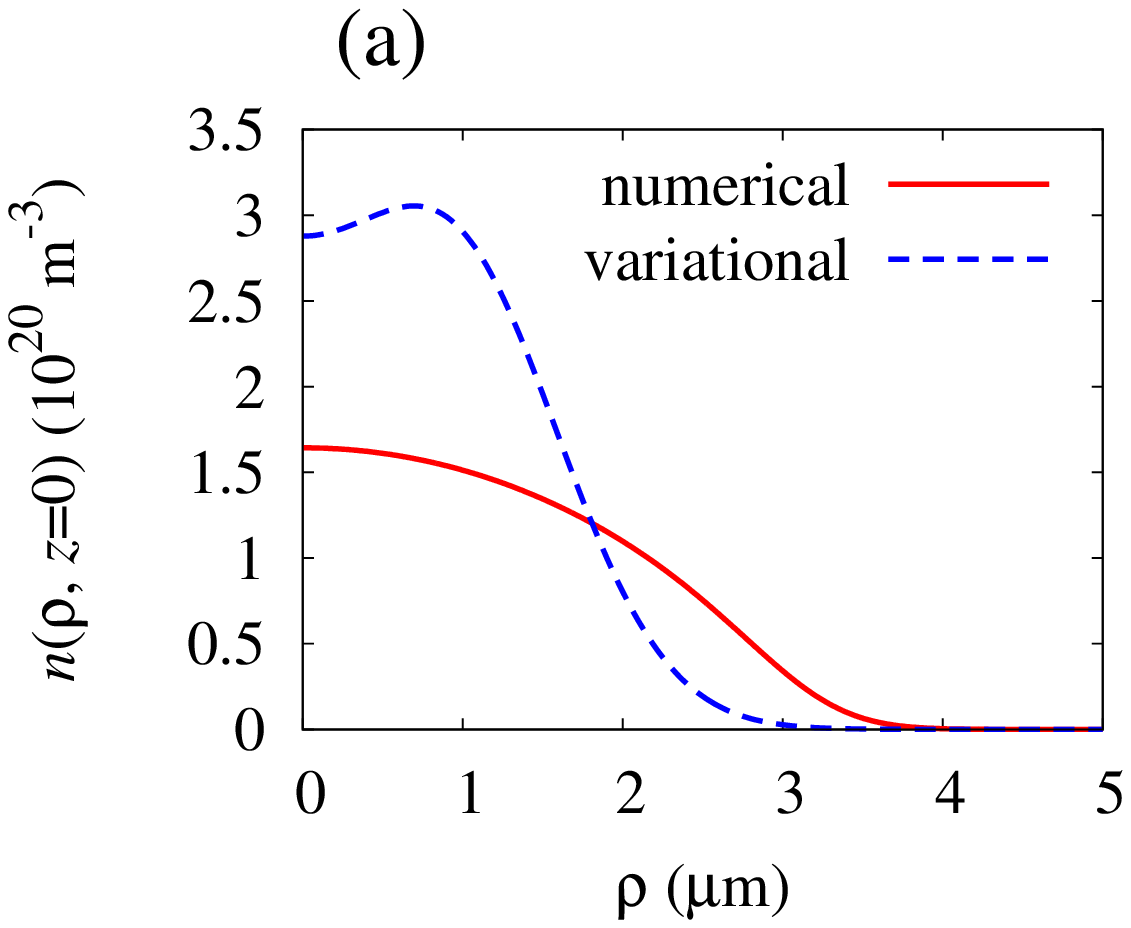}\hspace*{5mm}
\includegraphics[height=5.9cm]{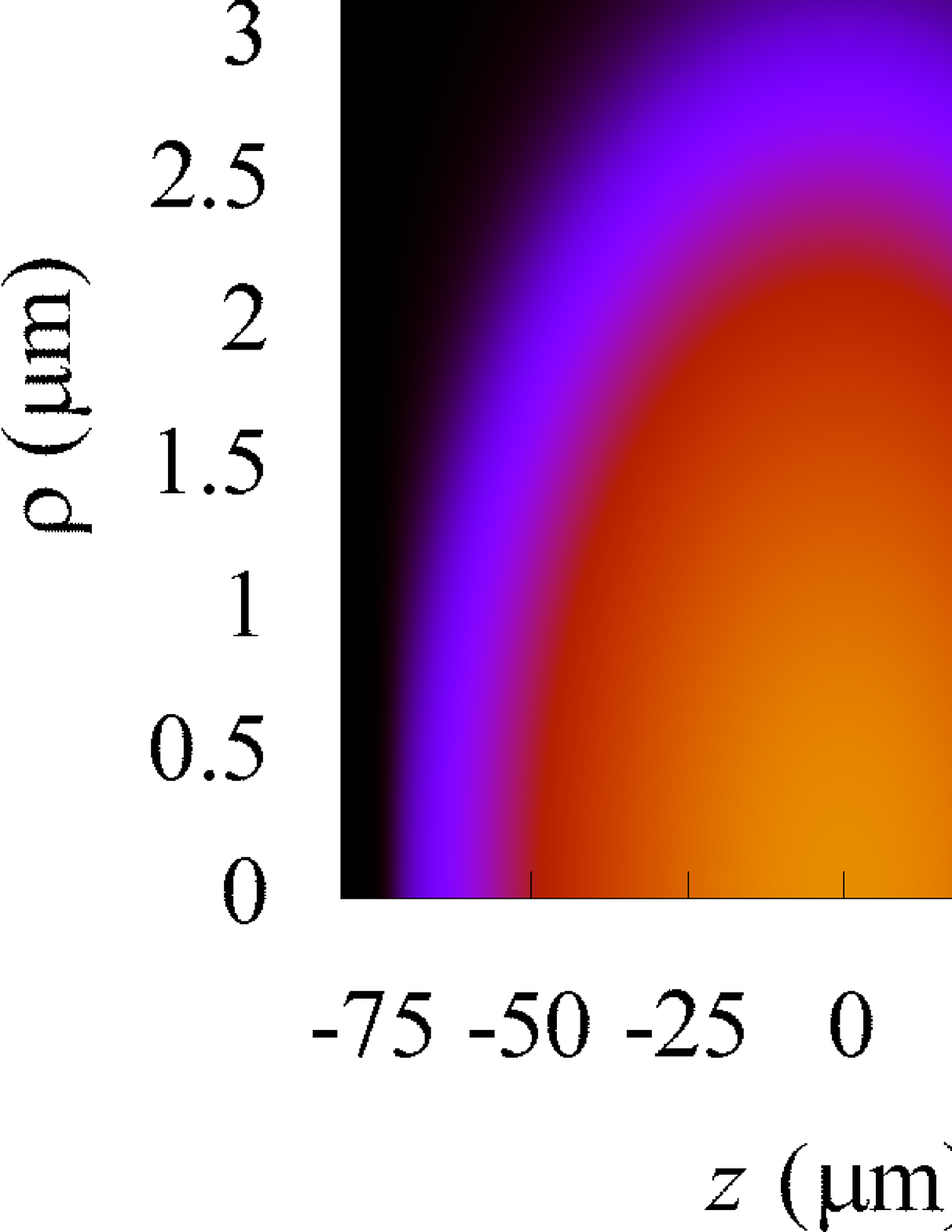}\\
\includegraphics[height=5.9cm]{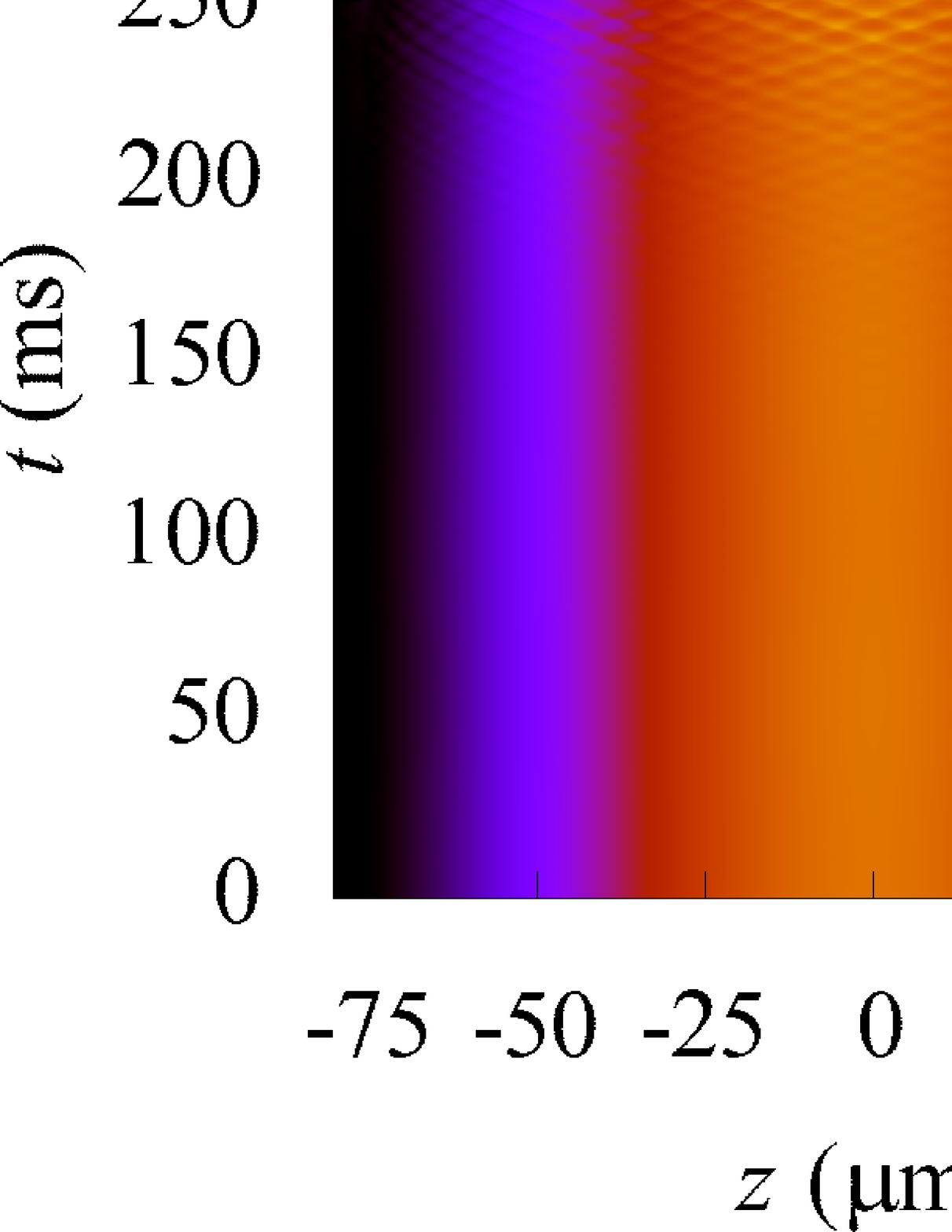}\hspace*{5mm}
\includegraphics[height=5.9cm]{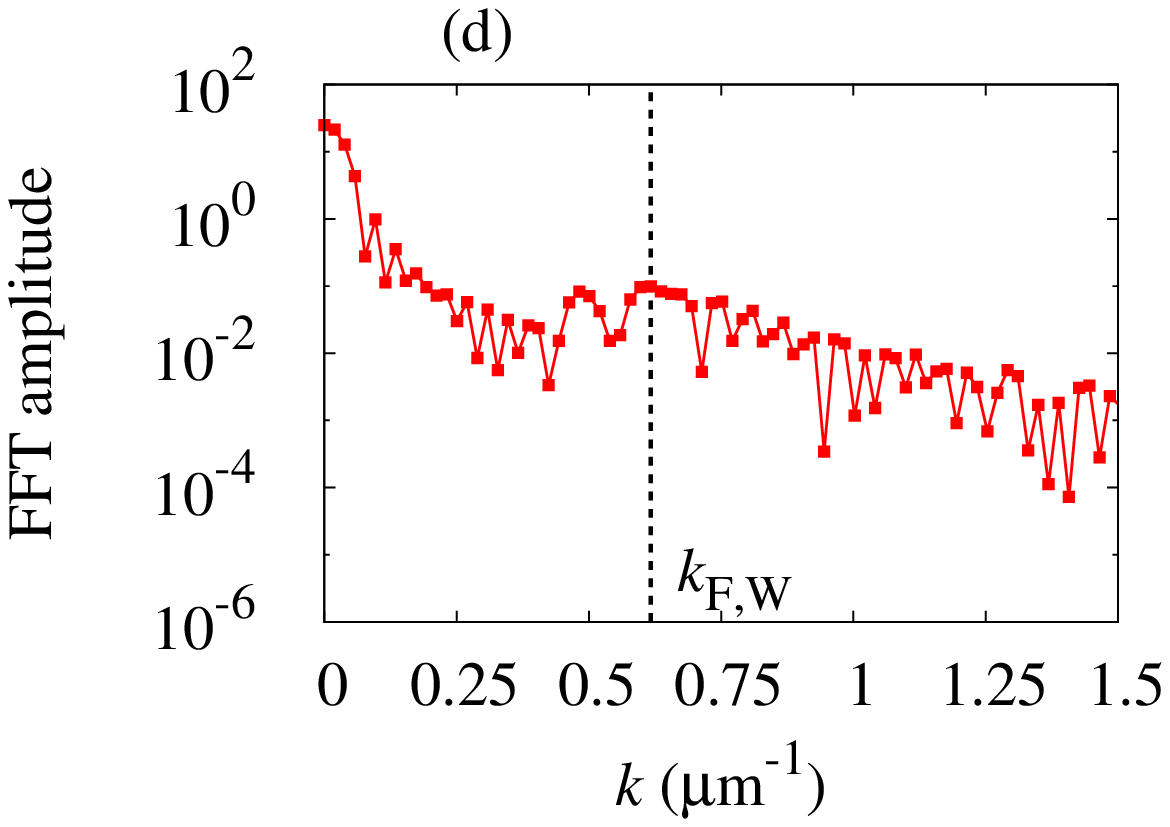}
\caption{(Color online) Weakly inhomogeneous collisions, inhomogeneity parameter $b=4b_0$.
(a) Radial component of the density profile at $z=0$ for the condensate ground state. The full red line shows the GPE numerical results, while the dashed blue line corresponds to the variational results.
(b) Full $\rho-z$ density profile of the ground state of the condensate.
(c) Time evolution of the radially-integrated longitudinal density profile
obtained with the modulation amplitude $\epsilon=0.1$ and the modulation frequency $\omega=250\times2\pi$~Hz.
The Faraday wave becomes fully visible after 200~ms.
(d) Fourier spectrum of the longitudinal density profile of the condensate at $t=250$~ms. The peak at $k_\mathrm{F,W}=0.60\, \mu$m$^{-1}$ corresponds to the Faraday wave, yielding a spatial period of $p=2\pi/k_\mathrm{F,W}=10.5\, \mu$m.}
\label{fig:fig1}
\end{figure*}

In this section we compare the variational results from the previous section with numerical results for a realistic
condensate with $N=2.5\times10^5$ atoms of $^{87}$Rb loaded into
a magnetic trap with frequencies $\Omega_{\rho 0}=160\times2\pi$~Hz and
$\Omega_{z}=7\times2\pi$~Hz.
First, by means of a standard imaginary time propagation, using the split-step semi-implicit Crank-Nicolson method \cite{GPENumerics2}, we determine the ground state of the condensate for the case of a constant scattering length $a_s=100.4\, a_0$,
and calculate the radial width of the condensate.
This radial width, hereafter designated $b_0$, is found to be $b_{0}=1.86\, \mu$m for given parameters of the system, and
serves as a referent length scale for values of inhomogeneity parameter $b$ in Eq.~(\ref{eq:g}).
Second, using the same imaginary-time propagation method, we determine the ground state of the condensate for a number of values of $b$, ranging from small ($b=b_0/4$) to large ($b=4b_0$), as well as for the limit of homogeneous interactions ($b\rightarrow\infty$). For each calculated ground states, we numerically simulate the real-time dynamics \cite{GPENumerics2} and monitor the emergence of Faraday patterns in the radially-integrated density profiles for parametric drives of the form $\Omega_{\rho}(t)=\Omega_{\rho 0}(1+\epsilon\sin\omega t)$,
where $\epsilon$ and $\omega$ represent modulation amplitude and frequency.

\begin{figure*}
\includegraphics[height=5.9cm]{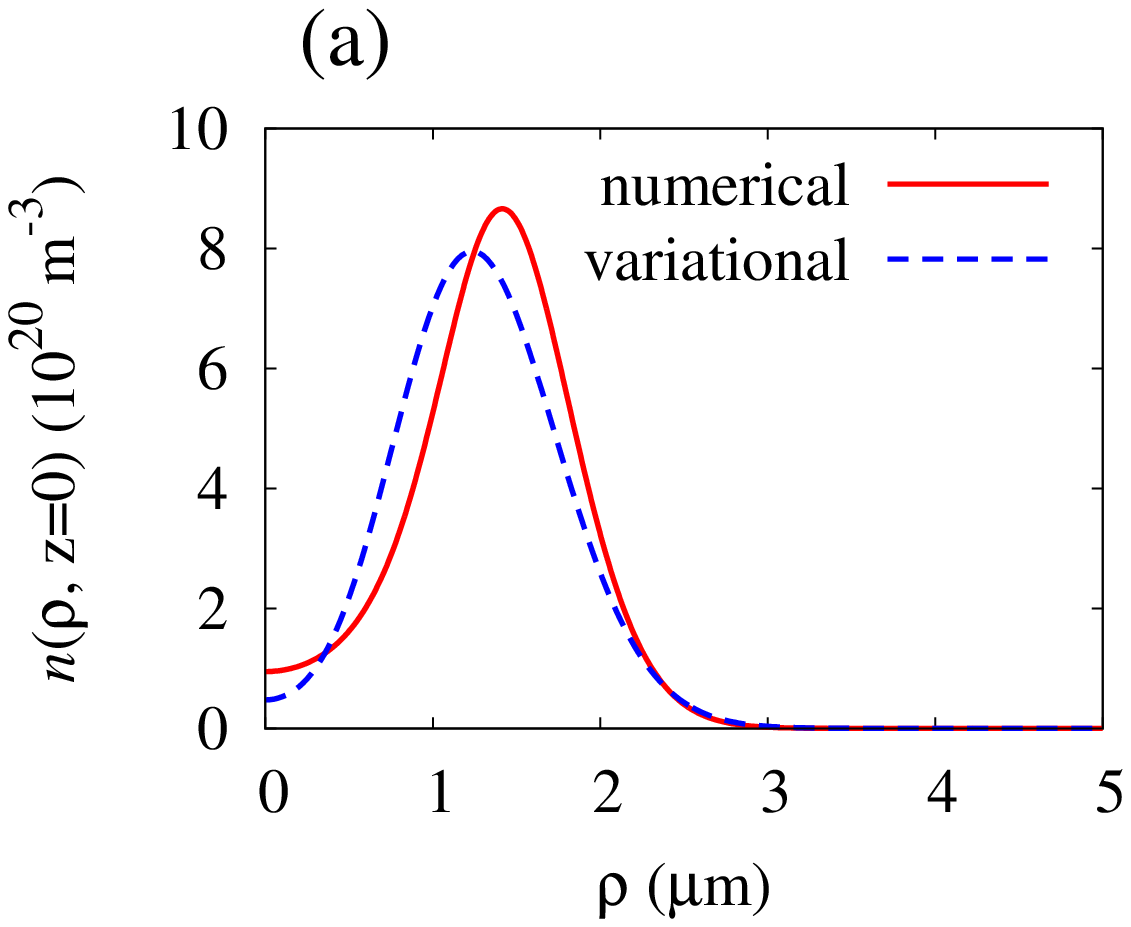}\hspace*{5mm}
\includegraphics[height=5.9cm]{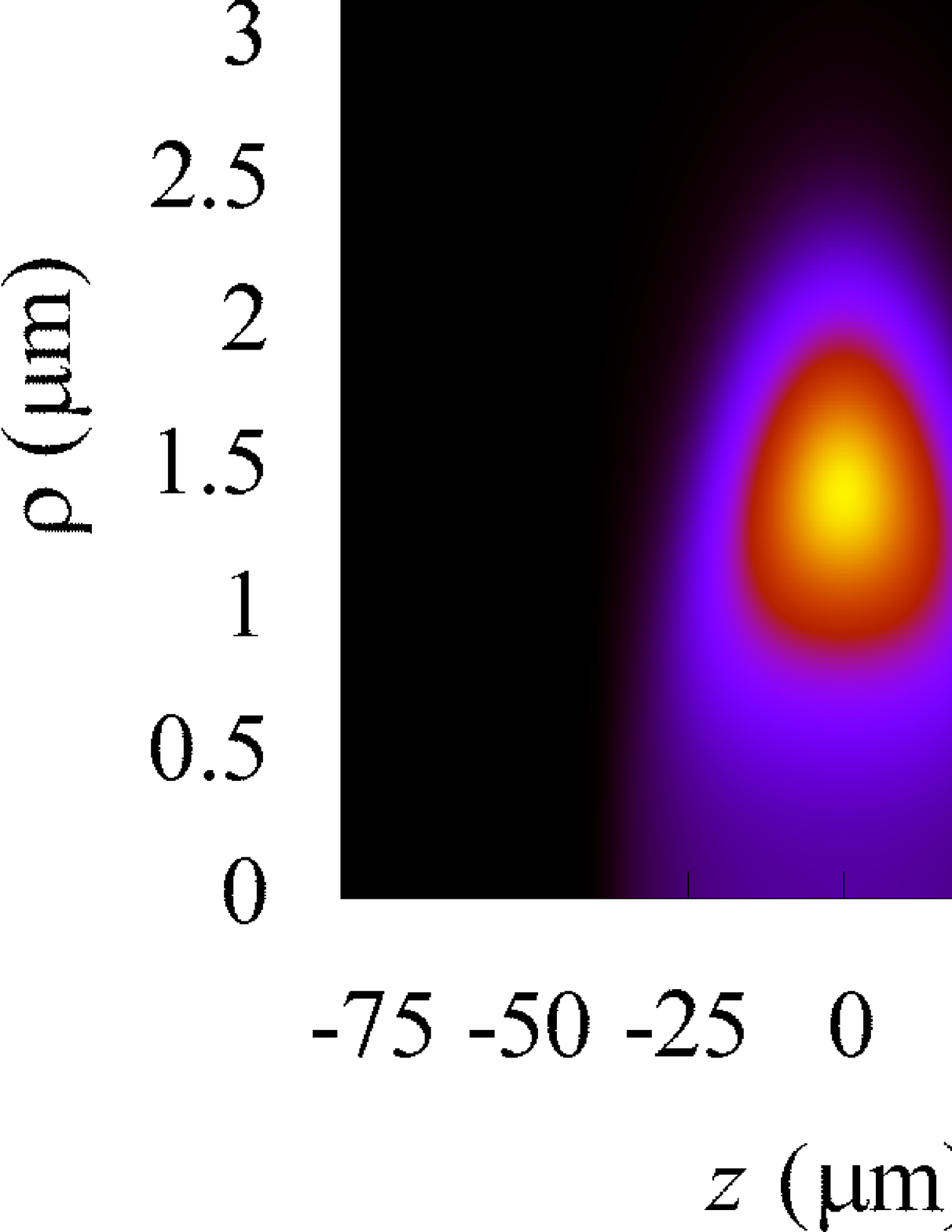}
\includegraphics[height=5.9cm]{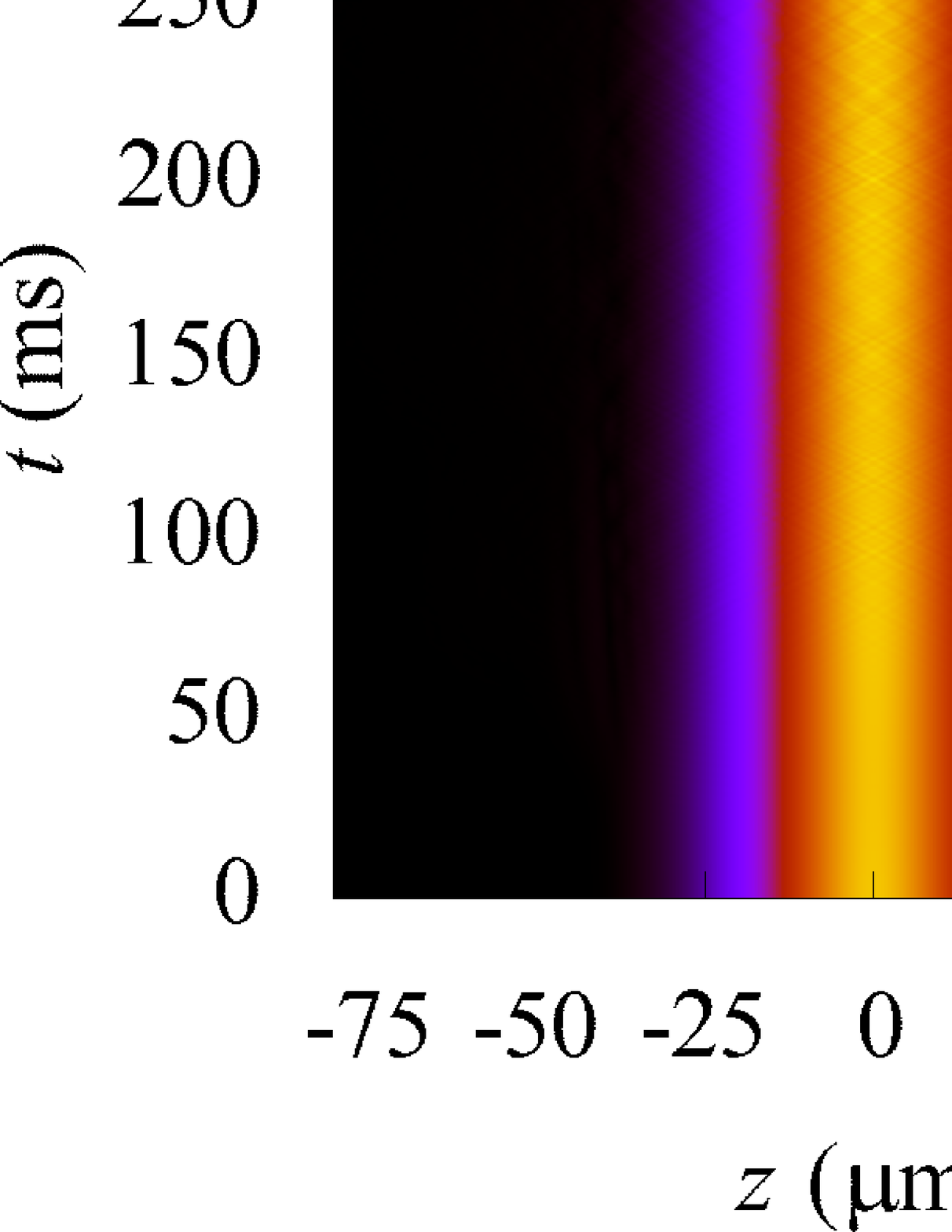}\hspace*{5mm}
\includegraphics[height=5.9cm]{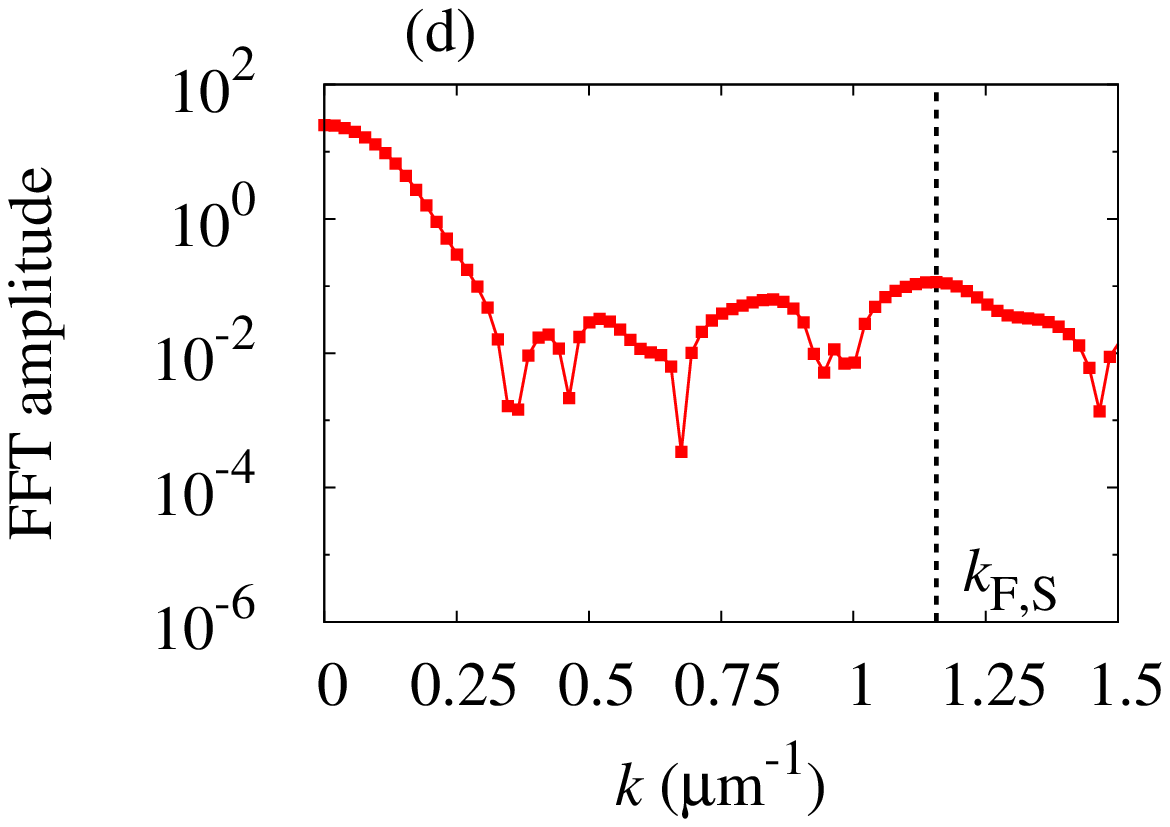}
\caption{(Color online) Strongly inhomogeneous collisions, inhomogeneity parameter $b=b_0/4$.
(a) Radial component of the density profile at $z=0$ for the condensate ground state. The full red line shows the GPE numerical results, while the dashed blue line corresponds to the variational results.
(b) Full $\rho-z$ density profile of the ground state of the condensate.
(c) Time evolution of the radially-integrated longitudinal density profile
obtained with the modulation amplitude $\epsilon=0.1$ and the modulation frequency $\omega=250\times2\pi$~Hz.
The Faraday wave becomes fully visible after 200~ms.
(d) Fourier spectrum of the longitudinal density profile of the condensate at $t=250$~ms. The peak at $k_\mathrm{F,S}=1.16\, \mu$m$^{-1}$ corresponds to the Faraday wave, yielding a spatial period of $p=2\pi/k_\mathrm{F,S}=5.4\, \mu$m.}
\label{fig:fig2}
\end{figure*}

In Figs.~\ref{fig:fig1}(a) and \ref{fig:fig1}(b) we show the radial density profile for $z=0$ and the full $\rho-z$ density profile of the ground state for $b=4b_0$, which corresponds to the case of weakly inhomogeneous collisions.
Fig.~\ref{fig:fig1}(c) gives the subsequent time evolution of the radially-integrated (column) density profile
of the condensate after modulation is switched on, with $\epsilon=0.1$ and $\omega=250\times2\pi$~Hz.
The panels (a) and (b) show a clear Thomas-Fermi density profile, while the variational result
obtained from the equilibrium solution of Eqs.~(\ref{eq:wicgamma})
and (\ref{eq:wicw}) significantly overestimates the peak density and underestimates
the radial extent of the condensate. Despite these quantitative differences,
we will see that the proposed ansatz captures the main features of the emergence of
density waves. These become fully visible around 200~ms after the start of modulation, as can be seen in Fig.~\ref{fig:fig1}(c).
In order to determine the spatial period of emerging Faraday patterns, in Fig.~\ref{fig:fig1}(d) we show the Fourier spectrum in the spatial domain of the radially-integrated
density profile at 250~ms after modulation is switched on. Note that due to the longitudinal component of
the magnetic trap the peaks in the Fourier spectrum always have finite
widths, which indicates the presence of a range of periods instead
of a single one.

This effect is even more pronounced in the case of strongly inhomogeneous
collisions (i.e., for small values of $b$), as we can see in Fig.~\ref{fig:fig2}(d).
This Fourier spectrum corresponds to the radially-integrated
density profile of a condensate for $b=b_0/4$, again calculated 250~ms after the modulation is switched on.
The full real-time dynamics of the condensate is depicted in Fig.~\ref{fig:fig2}(c),
where we clearly observe that such strongly inhomogeneous collisions in the radial direction
decrease the longitudinal extent of the condensate by a factor of two, as compared
to the case of weak inhomogeneity ($b=4b_{0}$) in Fig.~\ref{fig:fig1}(c). The redistribution of atoms in the condensate for the case of strong inhomogeneity is shown in Figs.~\ref{fig:fig2}(a) and \ref{fig:fig2}(b), where we immediately observe that the condensate has a radial density profile akin to that of a hollow cylinder.
 
\begin{figure}[!t]
\includegraphics[width=8.5cm]{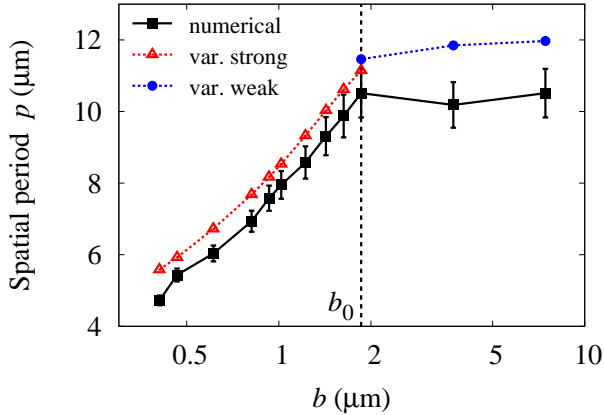}
\caption{(Color online) Average spatial period of the longitudinal Faraday waves as
a function of the inhomogeneity scale $b$ for the modulation amplitude $\epsilon=0.1$ and the modulation frequency
$\omega=250\times2\pi$~Hz. Black squares depict the full numerical
results obtained from the Fourier analysis of the solution of time-dependent GPE, red triangles and blue circles
correspond to the variational prediction of the spatial period obtained as
$2\pi/\bar{k}$, with $\bar{k}$ given by Eq.~(\ref{eq:averagedk}).
The red triangles are obtained using the dispersion 
relation (\ref{eq:dispersionB}) for strong inhomogeneity, while the blue circles are obtained from the 
dispersion relation (\ref{eq:dispersionA}) for weak inhomogeneity.}
\label{fig:fig3}
\end{figure}

When analyzed by Fourier transformation in the time domain, the density waves which appear in
Figs.~\ref{fig:fig1}(c) and \ref{fig:fig2}(c) have an intrinsic
frequency equal to half that of the drive $\omega$, and therefore can be identified as Faraday waves.
We have verified this numerically for all values of $b$ under scrutiny. In Fig.~\ref{fig:fig3} we show the spatial 
period of the observed Faraday waves for $\omega=250\times2\pi$~Hz as a function of inhomogeneity scale $b$ and compare the
numerical results with the analytic ones obtained in the previous
section.
Due to the finite widths of the
peaks in the Fourier spectra from which the spatial periods are determined numerically, we have associated an error bar to the average spatial period by taking the width of the dominant peak into account. The variational results shown in Fig.~\ref{fig:fig3} are obtained
from Eq.~(\ref{eq:dispersionA}) for weak inhomogeneity (blue circles, designated var.~weak) and from Eq.~(\ref{eq:dispersionB})
for strong inhomogeneity (red triangles, designated var.~strong), using an overall longitudinal Thomas-Fermi (TF) approximation which accounts
for the intrinsic longitudinal inhomogeneity of the density profile.

To this end, for $z\in [-L, L]$ we consider the longitudinal TF density profile of the form
\begin{equation}
n(z)=3\frac{L^{2}-z^{2}}{4L^{3}}\, ,\label{eq:triallongdensprofile}
\end{equation}
and $n(z)=0$ otherwise, and determine the average spatial 
period of the density wave from the corresponding wave vector defined
as 
\begin{equation}
\bar{k}=\frac{1}{2L}\int_{-L}^{L}k(z)\, dz\, .\label{eq:averagedk}
\end{equation}

Such kind of improvement has been already used to capture
quantitatively the dynamics of density waves in
cigar-shaped condensates of $^{87}$Rb \cite{FWHistF,VariationalDensityWavesF}.
The longitudinal TF extent of the condensate $2L$ is 
determined as follows. First, the wave function in the time-dependent GPE (\ref{eq:GPEtdep}) is assumed to have the simple separated form $\psi(\mathbf r, t)=\phi(\rho) f(z, t)$, where we have neglected all radial dynamics so that the function $\phi(\rho)$ is the stationary radial component of $\psi$, while the function $f(z, t)$ describes purely longitudinal dynamics. Second, the right-hand and the left-hand side of the time-dependent GPE is multiplied by $2\pi\rho\, \phi(\rho)$, and integrated over the radial coordinate $\rho$. This yields the one-dimensional time-dependent GPE
\begin{equation}
i\hbar\frac{\partial \tilde f}{\partial t}=\left(-\frac{\hbar^2}{2m}\frac{\partial^2}{\partial z^2}+\frac{1}{2}\Omega_z z^2+g_\mathrm{1D} N |\tilde f|^2\right)\tilde f\, ,
\label{eq:GPERtdep}
\end{equation}
where the effective one-dimensional interaction is
\begin{equation}
g_\mathrm{1D}=g_0\int_0^\infty d\rho\, 2\pi\rho\,  \phi(\rho)^4e^{-\frac{\rho^2}{2 b^2}}\, .
\end{equation}
and the function $f$ is rescaled by a phase factor so as to include contribution of integration of the radial component of the trapping potential, which only shifts the overall chemical potential.
In the third step, we apply the standard TF approximation for GPE (\ref{eq:GPERtdep}) and obtain the longitudinal extent of a BEC,
\begin{equation}
2L=\left(\frac{12 Ng_\mathrm{1D}}{\Omega_z^2}\right)^{1/3}\, ,
\end{equation}
where $g_\mathrm{1D}$ is computed using the stationary radial wave function obtained from Eqs.~(\ref{eq:wicgamma}) and (\ref{eq:wicw}) in the case of weakly inhomogeneous collisions, and Eqs.(\ref{eq:sicgamma}) and (\ref{eq:sicw}) for the strongly inhomogeneous case.

\begin{figure}[!t]
\includegraphics[width=7.1cm]{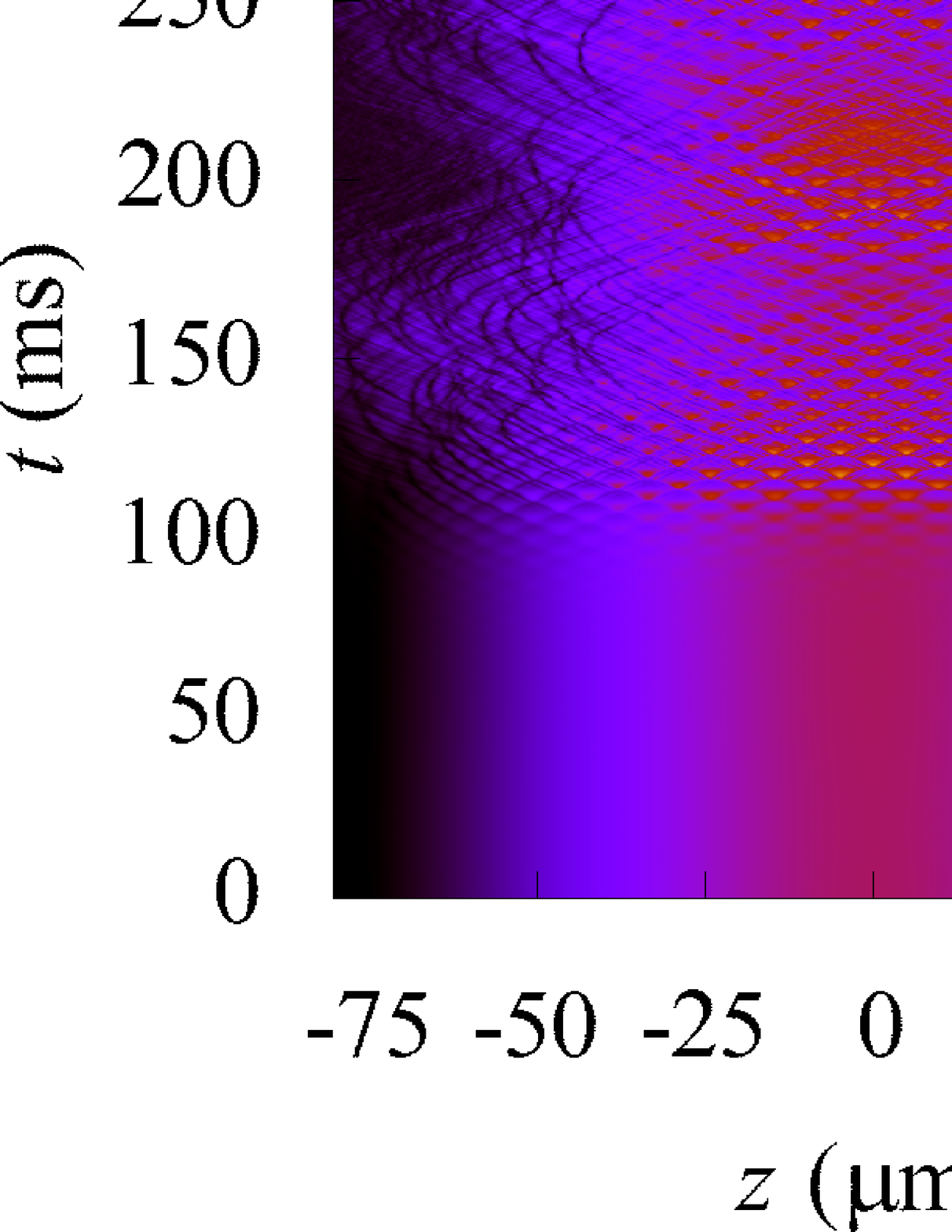}
\includegraphics[width=7.1cm]{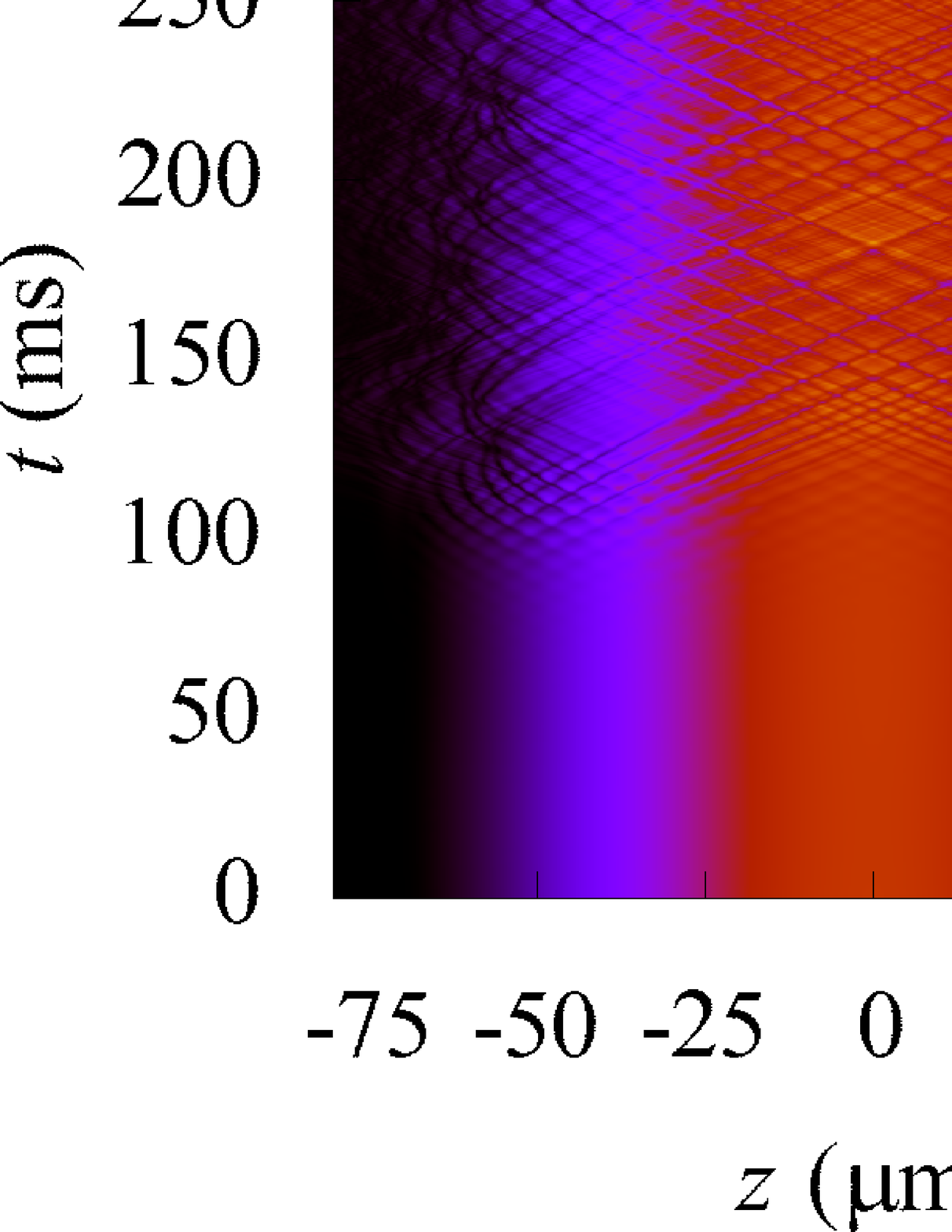}
\includegraphics[width=7.1cm]{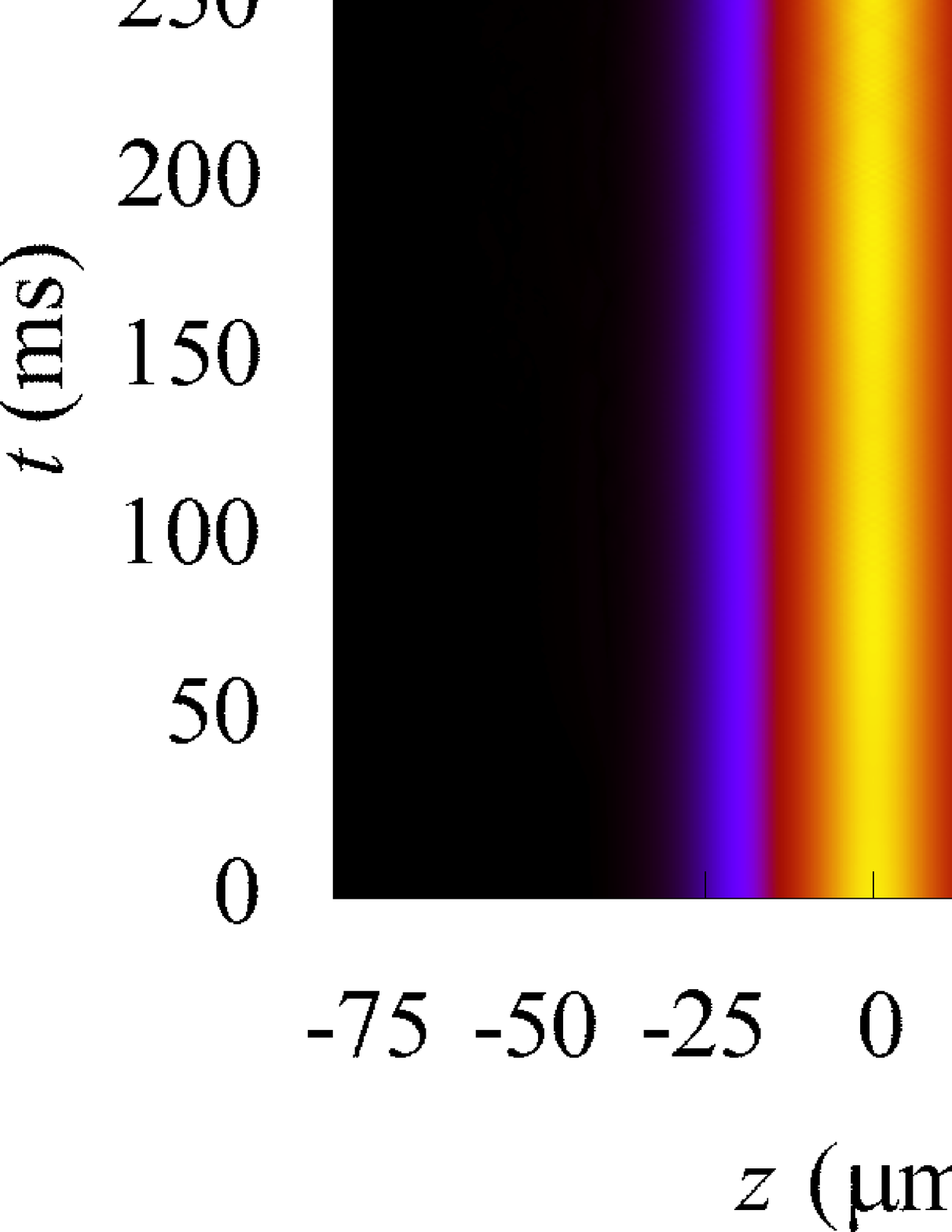}
\caption{(Color online) Time evolution of the radially-integrated longitudinal density profile
obtained with the modulation amplitude $\epsilon=0.1$ and the modulation frequency $\omega=160\times2\pi$~Hz for:
(a) $b=4b_0$; (b) $b=b_0$; (c) $b=b_0/4$. The excited collective modes soften for smaller values of $b$.}
\label{fig:fig4}
\end{figure}

In our numerical simulations we have seen Faraday waves for all non-resonant
drives (i.e., when the ratio $\omega/\Omega_{\rho 0}$ is not an integer number).
The resonant and near-resonant dynamics of the condensate
differs from the non-resonant one in two ways: first, the emergence
of density waves is accompanied by the excitation of a collective
mode (an effect which is particularly strong for weakly inhomogeneous
collisions) and second, the intrinsic frequency of the density wave
is equal to that of the drive, not half its value. In Fig.~\ref{fig:fig4} we illustrate
the resonant dynamics of a collisionally inhomogeneous condensate
for $b=4b_0$, $b=b_0$ and $b=b_0/4$ for a driving frequency
$\omega=\Omega_{\rho 0}=160\times2\pi$~Hz. The collective excitation is similar to that 
obtained experimentally by Pollack {\it et al.}~\cite{Pollack} in that both have the oscillations 
of the longitudinal extent of the condensate. However, in our case the radial
extent is roughly constant apart for small-amplitude oscillations triggered
by the periodic modulation of the radial component of the trap. Effectively,
we have a one-dimensional collective oscillation mode in the longitudinal
direction, while the dynamics of the radial extent is determined by the
external drive.

\begin{figure}[!t]
\includegraphics[width=7.1cm]{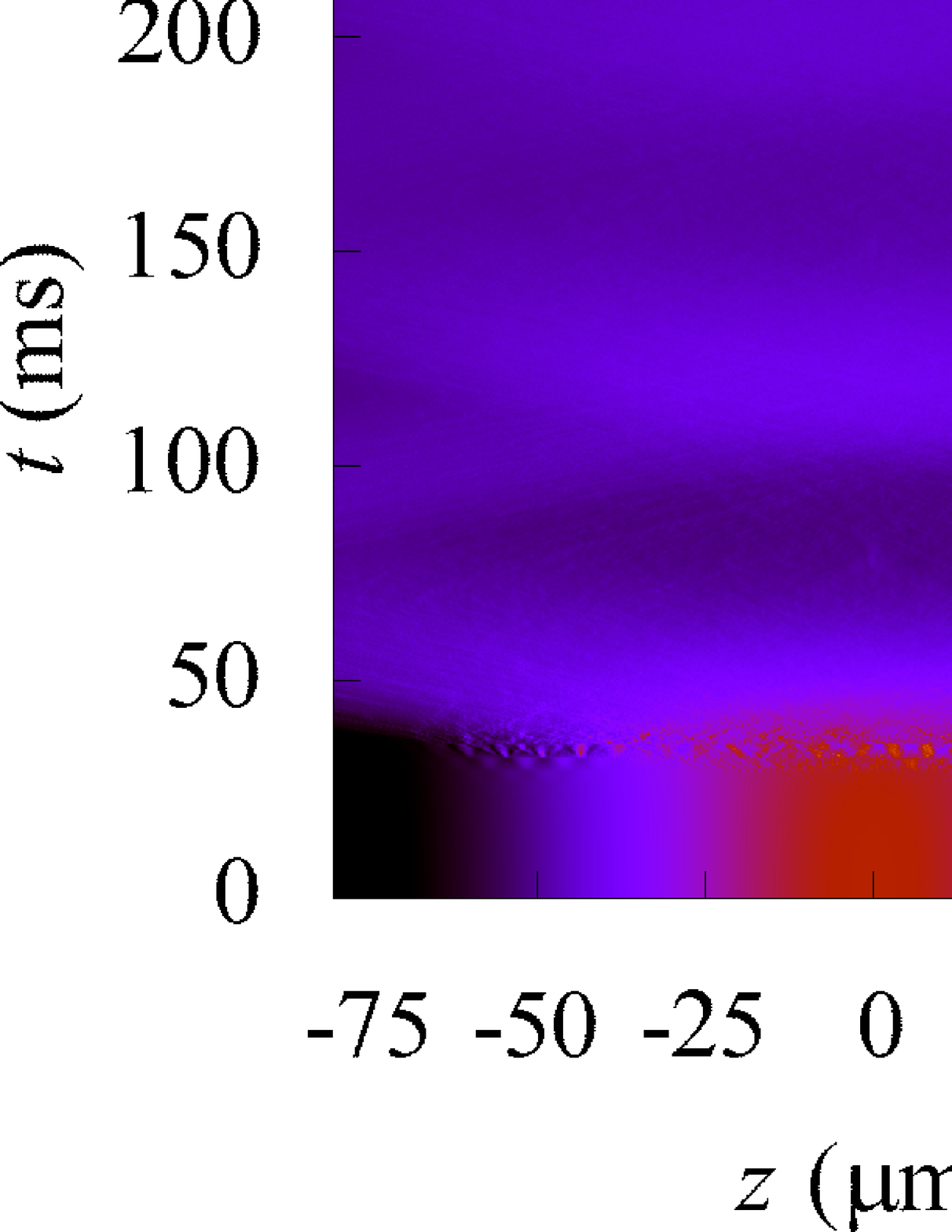}
\includegraphics[width=7.1cm]{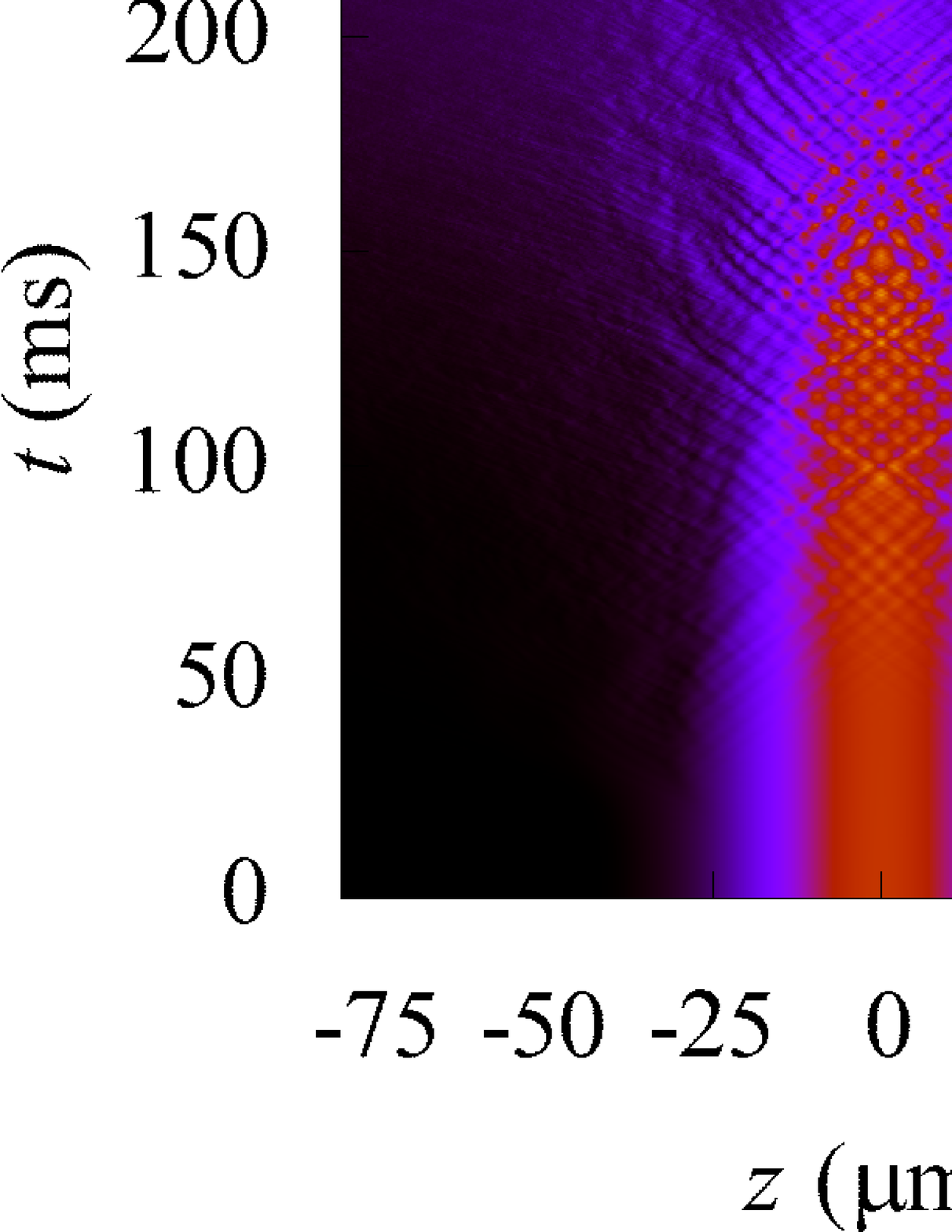}
\caption{(Color online) Time evolution of the radially-integrated longitudinal density profile
obtained with the modulation amplitude $\epsilon=0.1$ and the modulation frequency $\omega=320\times2\pi$~Hz for:
(a) $b=b_0$; (b) $b=b_0/4$.  In panel (a) the condensate destabilizes violently after the Faraday
wave sets in, while in panel (b) the destabilization is slower and one can clearly see the formation and subsequent evolution
of the Faraday wave.}
\label{fig:fig5}
\end{figure}

When comparing the dynamics for different values of inhomogeneity parameter $b$ in Fig.~\ref{fig:fig4}, the
main conclusion is that the collective dynamics softens for 
strongly inhomogeneous collisions (i.e., small values of $b$), which means that 
the instability onset times of the resonant waves increase as the inhomogeneity of
the collisions gets stronger and also that the amplitudes of the longitudinal oscillations get smaller. Note also that the localization
of the two-body collisions (and therefore the nonlinearity) close
to the symmetry axis of the system ($\rho=0$) effectively
turns the condensate into a linear system, so that nonlinear features such
as the aforementioned excitation of collective modes and density waves
are substantially slowed down. In fact, one can easily quantify the
effective nonlinearity by integrating out the radial component of
the interaction factor $g$. If we take into account Eq.~(\ref{eq:g}),
the resulting collisional factor is found to be 
\begin{equation}
g^*=\int_{0}^{\infty}d\rho\, 2\pi\rho\, g(\rho)=2\pi g_0 b^{2}\, ,
\label{eq:effectiveg}
\end{equation}
which shows that the effective nonlinear interaction depends quadratically on
$b$. This means that for strong inhomogeneity (small $b$) the bulk of the condensate
reaches an effectively linear regime (small $g^*$) in which nonlinear effects fade out.
The effective nonlinearity of the system increases with $b$ and so
does the spatial period of the Faraday waves, which are intrinsically
nonlinear waves. In other words, as the system departs from the linear
regime and becomes more nonlinear (increasing $b$), the Faraday
waves become more visible and have increasing spatial period, as well as
smaller instability onset time such that they can be easily identified.

One additional important conclusion that stems from results of our numerical simulations
is that there are substantial qualitative differences between the Faraday
waves which emerge at different driving frequencies. The spatial period of the wave
and its intrinsic frequency are enough to distinguish between the
Faraday waves excited by non-resonant drives. However, for driving frequencies
close to $\omega=2\Omega_{\rho 0}=320\times2\pi$~Hz (i.e., the second harmonic
of the radial frequency of the trap) this is not the case. The observed
waves have the spatial period and the frequency typical for Faraday waves,
but emerge more violently and have a dynamics similar to that of resonant
waves. This observation is particularly relevant for weakly inhomogeneous
collisions, as previous analytical and numerical studies of Faraday
waves focused chiefly on the spatial period and the frequency of the waves
and compared the theoretical results with the available experimental data from
Ref.~\cite{FWBECExp}. In Fig.~\ref{fig:fig5} we show the radially-integrated density profile for
$b=b_0$ and $b=b_0/4$. Note that the emergence of the Faraday
wave is accompanied by the excitation of a clear collective mode for
$b=b_0/4$, as we see in Fig.~\ref{fig:fig5}(b), while for $b=b_0$ the dynamics is
so forceful that, just after the wave sets in, the condensate quickly
becomes unstable, as can be seen in Fig.~\ref{fig:fig5}(a). In fact, we have observed
the same violent destabilization of the condensate after the Faraday wave
sets in all throughout the regime of weakly inhomogeneous collisions,
with almost no quantitative differences between $b=b_0$ and the limit of homogeneous interactions $b\rightarrow\infty$.

\section{Conclusions}
\label{sec:con}

We have studied the emergence of Faraday and resonant waves in cigar-shaped,
collisionally inhomogeneous Bose-Einstein condensates subject to periodic
modulations of the radial confinement. Using extensive
numerical simulations and detailed variational calculations, we have shown that for a Gaussian-shaped radially inhomogeneous scattering length the spatial period of the emerging Faraday waves increases as the inhomogeneity decreases,
and that it reaches a saturation plateau once the width of the Gaussian-shaped
inhomogeneity is close to the radial width of the condensate. The increase of the spatial period of the Faraday waves can be understood 
in terms of the effective nonlinearity of the system, which shows that the system becomes more nonlinear as the inhomogeneity becomes weaker, thereby exhibiting clearly observable Faraday waves of longer spatial periods and shorter instability onset times. 
Investigations
into the density profile of the condensate have shown that for strongly
inhomogeneous collisions the radial profile of the condensate is akin
to that of a hollow cylinder, while for the case of weak inhomogeneity the condensate is cigar-shaped and has a Thomas-Fermi radial density profile. Finally, we have shown that for modulation
frequencies close to the radial frequency of the trap the condensate
exhibits resonant waves accompanied by excitation of collective modes, while
for frequencies close to twice the radial frequency of the trap the
observed Faraday waves set in forcefully and are accompanied by energetic
collective modes which quickly destabilize the condensate for weakly
inhomogeneous collisions.

As a natural extension of this work, we plan to investigate the dynamics
of density waves excited through parametric resonance in cigar-shaped
condensates subject to thermal fluctuations. The interaction of the
condensate with the thermal cloud is particularly
relevant for long-timescale analysis, when the depletion of the condensate
due to the external drives can no longer be ignored.
We also plan to study two-dimensional
(pancake-shaped) condensates, which exhibit a rich variety
of density patterns, and where one could try to control the stability of patterns and
even generate spatio-temporal chaos by tuning the parametric drive (e.g., frequency and number of harmonics),
or the spatial patterns of inhomogeneous collisions (e.g., square or hexagonal), or both. Furthermore, we are planning to study the ramifications arising out of collisionally inhomogeneous interactions on Faraday waves in vector BECs.

\begin{acknowledgments}
For this work AB was supported in part by the Ministry of Education,
Science, and Technological Development of the Republic of Serbia under
projects ON171017 and NAI-DBEC, by DAAD - German Academic and
Exchange Service under project NAI-DBEC, and by the European Commission
under EU FP7 projects PRACE-2IP, PRACE-3IP, and EGI-InSPIRE.
AIN was supported by a grant of the Romanian Ministry of Education,
CNCS-UEFISCDI, under projects PN-II-RU-PD-2012-3-0145 and PN-II-ID-PCE-2011-3-0972.
SB wishes to acknowledge financial assistance from Department of
Science and Technology (Ref.~No.~SR/S2/HEP-26/2012) of the Government of India.
The work of RR forms a part of University Grants Commission (Ref.~No.~UGC-40-420/2011 (SR)),
Department of Atomic Energy - National Board of Higher Mathematics
(Ref.~No.~DAE-NBHM-2/ 48(1)2010/NBHM-RD II/4524), and Department of
Science and Technology (Ref.~No.~SR/S2/HEP-26/2012) of the Government of India.
\end{acknowledgments}

\end{document}